\documentclass[english,sorted&compressed]{aipproc}
\usepackage[T1]{fontenc}
\usepackage[latin1]{inputenc}
\usepackage{amsmath}
\usepackage{babel}
\usepackage{color}
\usepackage{graphics}
\usepackage{pifont}

\makeatletter

\providecommand{\LyX}{L\kern-.1667em\lower.25em\hbox{Y}\kern-.125emX\@}

 \layoutstyle{6x9}
 \renewcommand{\(}{$}
 \renewcommand{\)}{$}

\makeatother
\AtBeginDocument{

}

\begin{document}
\def\STAR{{\sc Star }}

\def\PHENIX{{\sc Phenix }}

\def\MSbar{\overline{MS}}

\title{Resummation for single-spin asymmetries in \( W \)~boson production}

\author{\underline{Pavel M. Nadolsky}}{address={Department of Physics, Southern Methodist University, Dallas, TX 75275-0175, U.S.A.},email={nadolsky@mail.physics.smu.edu}}

\author{C.-P. Yuan}{address={Department of Physics \& Astronomy, Michigan State University, East Lansing, MI 48824-2320, U.S.A.},email={yuan@pa.msu.edu}}

\date{12 October, 2002}

\copyrightyear{2002}

\begin{abstract}
To measure spin-dependent parton distribution functions in the production
of \( W^{\pm } \) bosons at the Relativistic Heavy Ion Collider, an accurate
model for distributions of charged leptons from the \( W \) boson decay
is needed. We present single-spin lepton-level cross sections of order \( {\cal O}(\alpha _{S}) \)
for this process, as well as resummed cross sections, which include multiple
parton radiation effects. We also present a program {\sc RhicBos} for the
numerical analysis of single- and double-spin cross sections in \( \gamma ^{*},\, W^{\pm }, \)
and \( Z^{0} \) boson production.
\end{abstract}
\maketitle
The measurement of longitudinal spin asymmetries in the production of \( W^{\pm } \)
bosons at the Relativistic Heavy Ion Collider (RHIC) will provide an essential
probe of spin-dependent quark distributions at high scales \( Q^{2} \) \cite{RHICOverview, Wboson1}.
At \( pp \) center-of-mass energy \( \sqrt{s}=500 \) GeV, about \( 1.3\times 10^{6} \)
\( W^{+} \) and \( W^{-} \) bosons will be produced by the time the integrated
luminosity reaches \( 800\mbox {\, pb}^{-1} \). Due to the parity violation
in the \( Wq\bar{q} \) coupling, this process permits non-vanishing single-spin
asymmetries \( A_{L}(\xi ) \), defined here a\textcolor{black}{s\begin{equation}
\label{ALxi}
A_{L}(\xi )\equiv \frac{\frac{d\sigma (p^{\rightarrow }p\rightarrow WX)}{d\xi }-\frac{d\sigma (p^{\leftarrow }p\rightarrow WX)}{d\xi }}{\frac{d\sigma (p^{\rightarrow }p\rightarrow WX)}{d\xi }+\frac{d\sigma (p^{\leftarrow }p\rightarrow WX)}{d\xi }},\mbox {\, where\, }\xi =y_{W},y_{l},p_{Tl},\dots \, .
\end{equation}
 Here \( y_{W} \) is the rapidity of the \( W \) boson; \( y_{l} \) and
\( p_{Tl} \) are the rapidity and transverse momentum of the charged lepton
from the \( W \) boson decay in the lab frame, respectively.} The \emph{lowest-order}
expression for the asymmetry \( A_{L}(y_{W}) \) with respect to the rapidity
\( y_{W} \) of the \( W \) boson is particularly simple if the absolute
value of \( y_{W} \) is large. In that case, \( A_{L}(y_{W}) \) reduces
to the ratio \( \Delta q(x)/q(x) \) of the polarized and unpolarized parton
distribution functions \cite{Wboson2}. Furthermore, \( A_{L}(\xi ) \) tests
the flavor dependence of quark polarizations.

\begin{figure}[tb]
{\centering \resizebox*{!}{0.38\textheight}{\includegraphics{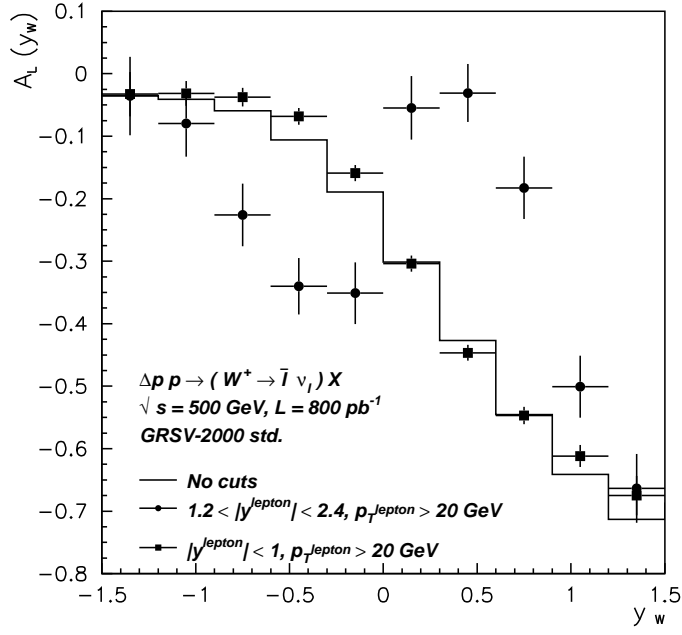}} \par}

\caption{\label{fig:ALyvscuts}Dependence of the asymmetry \protect\( A_{L}(y_{W})\protect \)
on the cuts imposed on the momentum of the observed antilepton in the \protect\( W^{+}\protect \)
boson production \protect\( p^{\rightarrow }p\rightarrow (W^{+}\rightarrow \bar{l}v_{l})X\protect \).
The asymmetry is calculated using the resummation method described in the
paper. The GRSV standard set \cite{Gluck:2000dy} of the polarized PDF's
was used. The error bars are calculated according to Eq.~(13) in Ref.~\cite{RHICOverview}
assuming the integrated luminosity \protect\( {\cal L}=800\mbox {\, pb}^{-1}\protect \)
and beam polarization \protect\( 70\%\protect \).}
\end{figure}

The original method for the measurement of \( W \) boson production at RHIC
is based on the direct reconstruction of the asymmetry \( A_{L}(y_{W}) \)
\cite{RHICOverview}. Unfortunately, such reconstruction is obstructed by
specifics of the detection of \( W^{\pm } \) bosons at RHIC. First, RHIC
detectors do not monitor energy balance in particle reactions. Due to the
lack of information about the missing momentum carried by the neutrino, the
determination of \( y_{W} \) is in general ambiguous and depends on assumptions
about the dynamics of the process. Second, due to the correlation between
the spins of the initial-state quarks and final-state leptons, the measured
value of \( A_{L}(y_{W}) \) is strongly sensitive to experimental cuts imposed
on the observed charged lepton. This feature is illustrated in Fig.~\ref{fig:ALyvscuts},
which shows the asymmetry in the \( W^{+} \) boson production calculated
without constraints on \( y_{l} \) and \( p_{Tl} \) (solid line), and with
constraints \( 1.2<|y_{l}|<2.4, \) \( p_{Tl}>20 \) GeV (circles) and \( |y_{l}|<1, \)
\( p_{Tl}>20 \) GeV (boxes). According to the Figure, there is a substantial
difference between the asymmetries calculated with and without selection
cuts. This difference arises due to the different dependence of the unpolarized
and polarized cross sections on angular distributions of the leptons in the
\( W \) boson decay. For instance, at the lowest order in \( W^{+} \)production\begin{equation}
A_{L}(y_{W},y_{l})=\frac{-\Delta u(x_{a})\bar{d}(x_{b})(1-\cos \theta ^{*})^{2}+\Delta \bar{d}(x_{a})u(x_{b})(1+\cos \theta ^{*})^{2}}{u(x_{a})\bar{d}(x_{b})(1-\cos \theta ^{*})^{2}+\bar{d}(x_{a})u(x_{b})(1+\cos \theta ^{*})^{2}},
\end{equation}
 where \( x_{a,b}\equiv (Q/\sqrt{s})e^{\pm y_{W}}, \) and \( \theta ^{*} \)
is the polar angle of the antilepton in the rest frame of the \( W \) boson.
Since \( y_{l} \) is related to \( \cos \theta ^{*}, \) as\[
y_{l}=y_{W}+\frac{1}{2}\ln \frac{1+\cos \theta ^{*}}{1-\cos \theta ^{*}},\]
 it is clear that restrictions on the range of integration of \( y_{l} \)
strongly affect \( A_{L}(y_{W}). \)

Since the only straightforward signature of the \( W \) bosons at RHIC is
the observation of secondary charged leptons, it is important to understand
differential cross sections of spin-dependent \( W \) boson production \emph{at
the lepton level}. Given that the radiative corrections are sizeable (\( \sim 30\% \))
both in the numerator and denominator of Eq.~(\ref{ALxi}), and that the
measurement results will be used in the next-to-leading order (NLO) PDF analysis,
it is necessary to derive these cross sections at NLO accuracy (\emph{i.e}.,
at order\nolinebreak~\( {\cal O}(\alpha _{S}) \)). 

Furthermore, most of the \( W \) bosons are produced with small, but non-zero
transverse momenta. Such non-zero \( q_{T} \) is acquired through radiation
of soft and collinear partons, which cannot be approximated by finite-order
perturbative calculations. In order to obtain reliable predictions for differential
cross sections, dominant logarithmic terms \( \alpha _{S}^{n}\ln ^{m}\left( q_{T}^{2}/Q^{2}\right)  \)
(where \( 0\leq m\leq 2n-1 \)) associated with such radiation should be
summed through all orders of the perturbative series. In our work \cite{NadolskyYuan2002},
we performed \textcolor{black}{a complete lepton-level study for the production
of \( W^{\pm },\, \gamma ^{*}, \) and \( Z^{0} \) bosons for arbitrary
longitudinal polarizations of incident protons. This study combined the \( {\cal O}(\alpha _{S}) \)
contributions with the all-order sum of small-\( q_{T} \) logarithmic corrections.}
The resummation of the logarithms \( \ln ^{m}(q_{T}^{2}/Q^{2}) \) was performed
with the help of the impact parameter space (\( b \)-space) resummation
technique \cite{CSS}. It extended the methodology developed for the unpolarized
vector boson production \cite{CPCsaba} to the spin-dependent case. 

Our study goes beyond those in the previous publications \cite{Kamal, Gehrmann, Weber}
in several aspects. It presents the fully differential NLO cross section
at the lepton level, which was not available before. The resummed single-
and double-spin cross sections for the production of \textit{on-shell} vector
bosons were presented earlier in Ref.\( \,  \)\cite{Weber}. We have derived
a more complete resummed cross section, which also accounts for the decays
of vector bosons, \emph{i.e.,} for spin correlations in the final state.
The lepton-level cross section includes several additional angular structure
functions, which do not contribute at the level of on-shell vector bosons.
Moreover, resummation is needed not only for the parity-conserving angular
function \( 1+\cos ^{2}\theta ^{*} \), which contributes to the on-shell
cross section, but also for the parity-violating angular function \( 2\cos \theta ^{*} \),
which affects angular distributions of the decay products. The estimate of
parity-violating contributions is more complicated since it involves \( \gamma _{5} \)-matrices
and Levi-Civita tensors both from the electroweak Lagrangian and spin-projection
operators. As a result, special care is needed to treat finite terms that
arise in the factorization of collinear poles in \( d\neq 4 \) dimensions.
We perform the calculation using the dimensional reduction and find that
the resummed cross sections satisfy helicity conservation conditions for
the incoming quarks. In addition, the coefficient functions in Ref.~\cite{Weber}
were obtained in a non-conventional factorization scheme and cannot be used
with the existing PDFs. In contrast, our results are fully consistent with
the \( \overline{MS} \) factorization scheme.

\begin{figure}[tb]
{\centering \resizebox*{1\columnwidth}{!}{\includegraphics{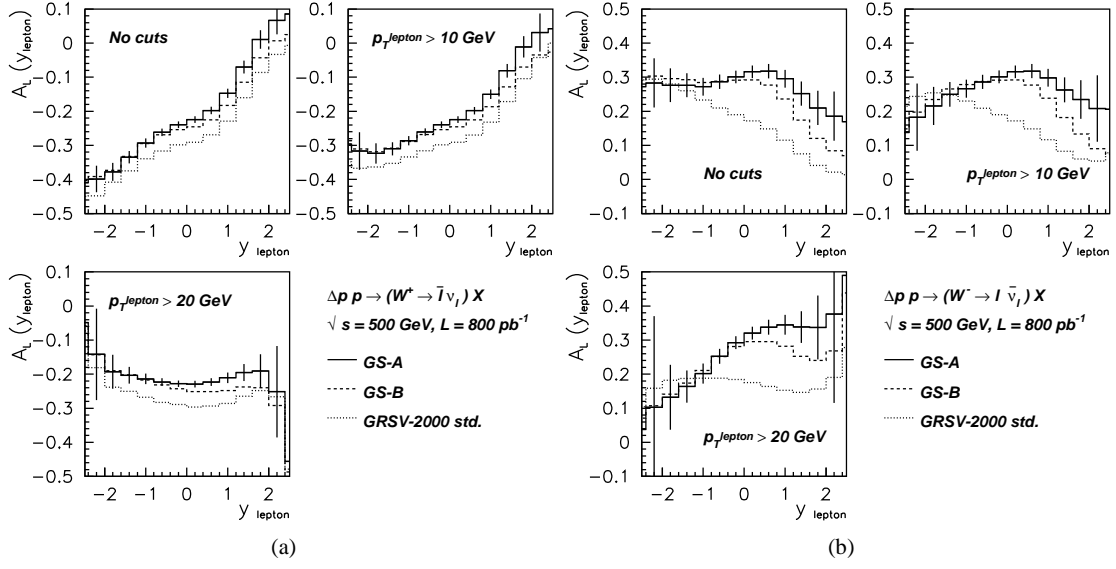}}  \par}

\caption{\label{fig:ALyl} Asymmetries \protect\( A_{L}(y_{l})\protect \) for various
selection cuts on \protect\( p_{Tl}\protect \) in (a) \protect\( W^{+}\protect \)
boson production and (b) \protect\( W^{-}\protect \) boson production. The
asymmetries are derived using the Gehrmann-Stirling PDF sets A and B \cite{GS}
and GRSV-2000 standard PDF set \cite{Gluck:2000dy}. Statistical errors are
estimated as in Fig.~\ref{fig:ALyvscuts}.}
\end{figure}

The resummed cross sections are incorporated in a numerical program {\sc RhicBos} for
Monte-Carlo integration of the differential cross sections \cite{RhicBos}.
We are not able to discuss all aspects of our numerical study in this short
report. However, as an example we show lepton-level asymmetries \( A_{L}(y_{l}) \)
for various cuts on \( p_{Tl} \) (Fig.~\ref{fig:ALyl}). We find that these
asymmetries can be accurately measured for both \( W^{+} \) and \( W^{-} \)
bosons. These directly observed asymmetries can efficiently discriminate
between different PDF sets; hence, they provide a viable alternative to the
less accessible asymmetry \( A_{L}(y_{W}) \). 

\begin{figure}[tb]
{\centering \resizebox*{!}{7.7cm}{\includegraphics{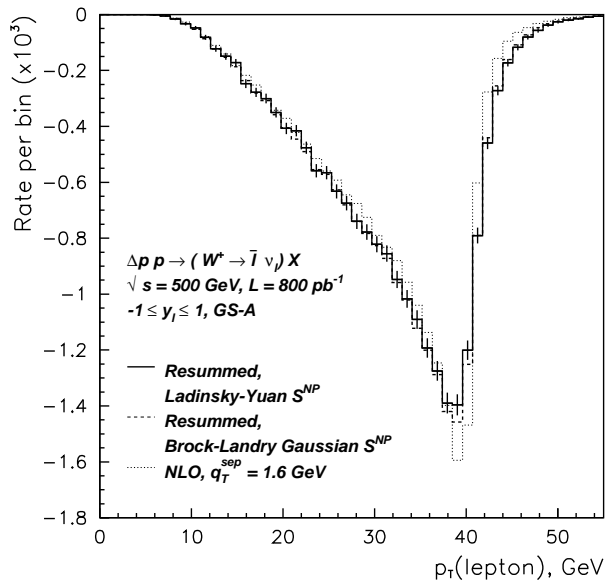}} \par}

\caption{\label{fig:ptl}The single-spin charged lepton transverse momentum distribution
for \protect\( W^{+}\protect \) boson production discussed in the main text.
The nonperturbative parts of the resummed cross sections were calculated
using the Ladinsky-Yuan parameterization \cite{LadinskyYuan} (solid) and
the most recent Brock-Landry Gaussian 2 parameterization \cite{BrockLandry}
(dashed).The \protect\( {\cal O}(\alpha _{S})\protect \) finite-order cross
section is shown as a dotted line.}
\end{figure}

It is also useful to study distributions with respect to the transverse momentum
\( p_{Tl} \) of the charged lepton, not only because they are sensitive
to the PDF's, but also because they probe in detail dynamics of the QCD radiation.
As was discussed above, the transverse momentum distributions for vector
bosons are affected by the multiple parton radiation, which can be described
only by means of all-order resummation. In addition, the distributions at
very small \( q_{T} \) are sensitive to nonperturbative contributions characterized
by large impact parameters \( b>1\mbox {\, GeV}^{-1} \). As a result, the
shape of the lepton-level distribution \( d\sigma /dp_{Tl} \) around its
peak at about \( p_{Tl}=M_{W}/2 \) is affected by both perturbative and
nonperturbative QCD radiation. Remarkably, the shape of the Jacobian peak
can be predicted by the theory, even though it cannot be calculated at any
finite order of \( \alpha _{S} \). The prediction is possible because the
perturbative soft and collinear contributions are systematically approximated
in the resummation formalism. The nonperturbative contributions currently
cannot be derived in a systematical way; but there is substantial indirect
evidence (spin independence of the perturbative soft radiation, quark helicity
conservation) that such contributions may be practically independent of the
proton spin and type of the vector boson. \textcolor{black}{}The impact of
the nonperturbative contributions is illustrated in Fig.~\ref{fig:ptl},
which shows the number of events for the difference \( d\sigma (p^{\rightarrow }p)/dp_{Tl}-d\sigma (p^{\leftarrow }p)/dp_{Tl} \)
of single-spin cross sections at \( {\cal L}=800\mbox {\, pb}^{-1} \). This
rate was calculated using two parameterizations \cite{LadinskyYuan, BrockLandry}
of the nonperturbative part, which were found in the unpolarized vector boson
production. It can be seen that the sensitivity to the nonperturbative input
is small, but, nonetheless, visible near the Jacobian peak. For comparison,
we also included the \( {\cal O}(\alpha _{S}) \) finite-order cross section
calculated using the phase space slicing method. The finite-order curve substantially
deviates from the resummed curves, and, moreover, its shape can be drastically
modified by varying the phase space slicing parameter \( q_{T}^{sep} \).
In contrast, the resummed curve is determined unambiguously once a parameterization
of the nonperturbative input is obtained from the double-spin \( \gamma ^{*} \)
production, single-spin or double-spin \( Z^{0} \) production, or even unpolarized
\( W \) production. Needless to say, the resummation predictions for the
shape of the Jacobian peak, which directly follow from fundamental principles
of QCD, must be tested at RHIC. 

\begin{theacknowledgments}
~The research of P. M. N. was supported by the U.S. Department of Energy,
National Science Foundation, and Lightner-Sams Foundation. The research of
C.-P. Y. was supported by the National Science Foundation under grant PHY-0100677.
\end{theacknowledgments}

\end{document}